\documentclass[5p,authoryear]{elsarticle}

\usepackage{mathrsfs}
\usepackage{graphicx}
\usepackage{amsmath}
\usepackage{amssymb}
\usepackage{amsfonts}
\usepackage{natbib}
\usepackage{hyperref}
\hypersetup{colorlinks=true,
citecolor=green,
linkcolor=red,
urlcolor=cyan,
pdftex}

\newcommand{\sss}{\scriptscriptstyle}
\newcommand{\mbf}{\mathbf}
\newcommand{\mrm}{\mathrm}
\newcommand{\ul}{\underline}

\newcommand{\ds}{\displaystyle}
\newcommand{\ovl}{\overline}
\newcommand{\eq}{equation}

\newcommand{\lra}{\leftrightarrow}

\begin{document}

\title{Inferring population statistics of receptor neurons sensitivities \\ and firing-rates from general functional requirements}

\author{Carlo Fulvi Mari\fnref{1}}
\fntext[1]{E-mail address: \href{mailto:cfmphys@gmail.com}{cfmphys@gmail.com} \\
\href{https://www.orcid.org/0000-0002-5828-9412}{https://www.orcid.org/0000-0002-5828-9412}} 
\date{Submitted 7 February 2020 -- Revised 7 April 2020}

\begin{abstract}
On the basis of the evident ability of neuronal olfactory systems to evaluate the intensity of an odorous stimulus and at the same time also recognise the identity of the odorant over a large range of concentrations, a few biologically-realistic hypotheses on some of the underlying neural processes are made. In particular, it is assumed that the receptor neurons mean firing-rate scale monotonically with odorant intensity, and that the receptor sensitivities range widely across odorants and receptor neurons hence leading to highly distributed representations of the stimuli. The mathematical implementation of the phenomenological postulates allows for inferring explicit functional relationships between some measurable quantities. It results that both the dependence of the mean firing-rate on odorant concentration and the statistical distribution of receptor sensitivity across the neuronal population are power-laws, whose respective exponents are in an arithmetic, testable relationship.

In order to test quantitatively the prediction of power-law dependence of population mean firing-rate on odorant concentration, a probabilistic model is created to extract information from data available in the experimental literature. The values of the free parameters of the model are estimated by an info-geometric Bayesian maximum-likelihood inference which keeps into account the prior distribution of the parameters. The eventual goodness of fit is quantified by means of a distribution-independent test.

The probabilistic model results to be accurate with high statistical significance, thus confirming the theoretical prediction of a power-law dependence on odorant concentration. The experimental data available about the distribution of sensitivities also agree with the other predictions, though they are not statistically sufficient for a very stringent verification. Furthermore, the theory suggests a potential evolutionary reason for the exponent of the sensitivity power-law to be significantly different from the unit. The power-law dependence on concentration is consistent with the psychophysical Stevens Law.

On the whole, from the formalisation of just a few phenomenological observations a compact model is derived that may fit experimental findings from several levels of research on olfaction.
\end{abstract} 

\begin{keyword}
Neural coding \sep Distributed representation \sep Olfaction \sep Receptor sensitivity \sep Affinity distribution \sep Concentration invariance \sep Power-law. \\
\vspace{.3cm}
DOI: \href{https://doi.org/10.1016/j.biosystems.2020.104153}{10.1016/j.biosystems.2020.104153} \hfill Ref.: C. Fulvi Mari, {\it BioSystems} (2020) 104153.\\
\vspace{.2cm}
\scriptsize{\copyright 2020. This manuscript version is made available under the CC-BY-NC-ND 4.0 licence (\href{https://creativecommons.org/licenses/by-nc-nd/4.0/}{https://creativecommons.org/licenses/by-nc-nd/4.0/}).}
\end{keyword}

\maketitle

\section{Introduction}
The sensory system of an animal must generally be able to evaluate the intensity of a stimulus as well as to recognise its identity, or possibly determine that it is not among those already learned, for reasons of survival of the individual or of the species (e.g., food search by odorant concentration gradient and friend/foe discrimination). This ability is in fact common to all sensory modalities of most animal species, including humans, and has long been of interest to psychophysics and neurophysiology, but determining its functional underpinnings has been a challenging task. In these regards, the olfactory system lends itself to the most extensive and accurate investigations because of the considerable simplicity and consistency of its architecture and basic functions across many species.

In the olfactory systems of insects and vertebrates the processing stages are compartmentalised and well localised in a few small regions, the connectivity between them is quite simple, and their neuronal populations are relatively easy to access in electrophysiological and fluorescence imaging experiments, also \emph{in vivo} and simultaneous. Additionally, the phenotype of the olfactory system, especially in its front end, that is, the olfactory receptor neurons, is determined by the genotype in such a direct way that the system is very amenable to genetic manipulations as in, for example, gene knockout and optogenetics experiments. All this, together with advances in physiology methods and technologies, has allowed for detailed experimental studies not only on the structure of the subsystems but also on their respective functions, in particular concerning the encoding of stimuli by the neuronal populations.

The main focus of the present article is on the statistical properties of the olfactory system front end, namely receptors and receptor neurons, and on the neural coding of odour intensity. It will be shown how both the statistical distribution of receptor affinity and the dependence of receptor neurons mean firing-rate on odorant intensity can be mathematically derived from just three simple phenomenological assumptions. The predictions of the mathematical theory will be verified against experimental data that are already available on highly reputable publications and then used to propose directions of further experimental investigations, also involving evolutionary elements.

\section{Essentials of an olfactory system}\label{essentials}
The basic functional anatomy of olfactory systems is quite simple and largely shared across very many species, from insects to mammals. In this Section, only a generalised qualitative description is presented, focusing on the main common features, and therefore it should only be considered as minimal information for the justification of the model assumptions; much more information is provided in the cited references and reviews \citep{Sachse2018,Su,Hildebrand} along with the references therein, the relevant literature being very conspicuous. An effort, likely only partially successful, was also made to cite the precursors of the experimental or theoretical work here explicitly instanced, including those that have been historically followed by several other publications of ever increasing accuracy and sophistication as further developments.

The front end of a typical olfactory system is made of \emph{olfactory receptor neurons} (ORNs), whose cilia are more or less directly exposed to the flow of the medium in which the animal lives. Each cilium expresses \emph{olfactory receptors} (ORs) to which the odorant molecules in the medium may bind. When a ligand binds with a receptor, a cascade of biochemical reactions inside the ORN is triggered that leads to the opening of membrane ion channels. As a consequence, influx and efflux of various types of ions take place, hence altering the difference of electric potential between the inside and the outside of the cell, which at rest is negative. In most cases the effect is one of depolarisation (excitatory); if the relative magnitude of the electric potential is brought up to a certain threshold, an action potential, or \emph{spike}, is generated that propagates forward through the axon of the ORN to downstream neural stages. In a minority of cases, the effect of the binding is one of hyperpolarisation (inhibitory), so that the usually present spontaneous activity of the ORN is reduced or even completely suppressed. The activation cascade is immediately followed by a negative feedback sequence that quickly compensates for the consequences of the former \citep{Kurahashi1990,Firestein1990,Zufall}. 

The affinity\footnote{The ORNs cilia are usually immersed in a watery layer; therefore, the concentration that matters is the one of the odorant molecules that, after absorption and diffusion, are present within such mucus rather than the concentration in the carrier medium. When measuring the concentration in the medium, affinity should be intended as an effective affinity.} between any OR and odorant may differ greatly across ORs and odorants. It has been clearly shown that generally each ORN expresses one only type of OR on its cilium, directly defined by a gene, and that several types of ORs are expressed across the population of ORNs \citep{Buck1991}. Because of all this, any odorant will normally evoke a pattern of \emph{distributed graded activity} across the population of ORNs \citep{Malnic1999,Rubin1999,Duchamp1999}, therefore providing a means to produce a different representation for any of a large number of odorants over a wide range of concentrations. 

Each ORN projects its axon into only one of the downstream \emph{glomeruli} that lie within the olfactory bulb in vertebrates or the antennal lobe in insects. The axons of the ORNs of the same class, that is, of the ORNs that express the same OR type, converge into the same glomeruli, and each glomerulus only receives afferents from ORNs of the same class. Each glomerulus is innervated by the apical dendrite of a \emph{relay neuron}, which does not innervate any other glomerulus, so that the pattern of spiking activity of the relay neurons faithfully reflects the ORs activation, then conveying information about the odorant to the central nervous system.

When the population of ORNs is exposed to an odorant, a pattern of spiking activity is evoked. As the concentration of the odorant pulses is increased, already active ORNs increase their respective firing-rates (FRs), unless they are already at saturation, and more and more ORNs are recruited as they are activated to the point of generating spikes. This of course changes the population output pattern \citep{Ma2000,Stopfer2003,Wachowiak2002} and therefore some kind of mechanism for \emph{concentration invariance} has to act in a downstream system in order to maintain the identity of the odorant, at least for an ecologically relevant range. The simplest way that may be conceived to normalise concentration dependence is naturally some form of divisive inhibition whose amplitude scale monotonically with the mean FR of the afferent pattern. Indeed, it was discovered that normalising divisive inhibition of relay neurons exists in the antennal lobe of insects \citep{Olsen2010}; as the structure of the glomerular network in vertebrates is anatomically and functionally very similar to that of insects, it seems most likely that divisive normalisation takes place there too. It is also known that at least some odorants may be identified differently if they are presented at very different concentrations, which hints to limitations in the normalisation mechanism.

The fact that the number of spiking ORNs also may change as the concentration of the odorant varies presents a further hurdle for the stability of odorant identification. This problem could be at least partly solved by exploiting pattern completion abilities of downstream recurrent neuronal networks \citep{Little1974,Hopfield1982,Amit1985,Treves1990b}. Segmentation of mixtures of odorants is also thought to be achieved by downstream systems, possibly by exploiting intrinsic dynamical instabilities of recurrent neuronal networks \citep{Marlsburg,Tsodyks1998}. However, such topics, still debated, are beyond the scope of the present work and will not be discussed further here.

\section{Model, analysis and predictions}
Because the odorant-induced spiking of any ORN is quickly reduced or suppressed by the intracellular negative feedback, the spikes considered in this work are only those generated within the first 500ms after the stimulus delivery, that is, before the negative feedback becomes relevant. Also, because of the $OR \ type \lra ORN \ class \lra relay \ neuron$ correspondence, from this point onwards in this article it will be conventionally assumed that the OR of each type is expressed by one and only one ORN, so that the total number of ORNs is equal to the total number of OR types (or, more simply, ORs). The sensitivity of any ORN to any odorant molecule will then be synonymic to the OR affinity to that same molecule, unless explicitly stated otherwise. As it will appear in the following, this simplification will cause no loss of generality.

The mean FR across a population of ORNs for any given odorant at concentration $c$ is defined by
\begin{equation}
\ovl{\nu}(c) \doteq \frac{1}{M} \sum_{i=1}^{M} \nu_{i}(c),
\end{equation}
where $M$ is the number of ORNs and $\nu_{i}$ is the FR of ORN $i$. Having assumed that each ORN expresses a different OR, every ORN can be labelled by its sensitivity to the presented odorant. Therefore, the FR of the ORN with sensitivity $K$ to the odorant that is presented at concentration $c$ is here indicated with $\nu(K,c)$.

Given an odorant and its concentration, because of the existence of the threshold voltage for neuronal action-potential generation, only the ORNs with sufficiently large sensitivity to that odorant will respond to the stimulus. The value of the sensitivity that corresponds to such threshold is a function of the odorant concentration here indicated with $\widehat{K}(c)$. For $M$ sufficiently large, postulating that on such large scale the possible statistical dependence between sensitivities is of minor importance, one can make the approximation
\begin{equation} \label{nuint}
\ovl{\nu}(c) \simeq \int_{\widehat{K}(c)}^{+\infty} \!\! dx \, f_{_{\!K}}\!(x) \, \nu(x, c),
\end{equation}
where $f_{_{\!K}}$ is the probability density function (PDF) of the sensitivity. 

Information on the concentration of an odorant is conveyed by the mean FR of the ORNs and, as described in Section \ref{essentials}, the downstream systems must also be able to compensate for changes of concentration in order to maintain a stable representation of the identity of the odorant. In view of the feasibility of normalisation mechanisms as, e.g., divisive inhibition, it seems reasonable to postulate that the mean FR scales monotonically with the odorant concentration:
\begin{equation} \label{nubar}
\ovl{\nu}(\eta \, c) = g(\eta) \, \ovl{\nu}(c),
\end{equation}
where $\eta \in \mathbb{R}^{+}$, $g$ is a differentiable monotonically increasing function on $\mathbb{R^{+}}$, and naturally $g(1)=1$. 

The third and last postulate requires that the FR of any ORN is a function of the sensitivity $K$ and of the odorant concentration $c$ through their product $K c$ only, that is (with some abuse of notation)
\begin{equation} \label{nuKc}
\nu(K,c)=\nu(K c).
\end{equation}
This assumption would be certainly correct if the stimulus-evoked cross-membrane current $I$ was proportional to the number of ligand-receptor bindings as derived from the Mass Action Law (MAL) of the reaction $L+R \rightleftarrows LR$, where $L$ and $R$ represent, respectively, a ligand (odorant molecule) and a receptor on the ORN cilium. However, it has been known for long \citep{Hill,Monod} that in biological systems the binding of a ligand to a receptor may modulate (e.g., allosterically) the probability that another ligand bind to the same receptor or one close to it \citep{Prinz}. The (statistical) consequence of this complex of phenomena at the molecular level is reflected into the Hill coefficient $\gamma$ in the Hill formula $I \propto 1/[1+(K c)^{-\gamma}]$ not being equal to 1, as it would be for the simple MAL, or any other positive integer, as it would be for the MAL of $\gamma L + R \rightleftarrows L_{\gamma} R$. In fact, the Ca$^{\sss 2+}$ influx current of the ORNs in the experiments by \citet{Samuel2019} was shown by the authors to be very well described by a Hill function with $\gamma \simeq 1.42$. The hypothesis that the dependence be on the product $K c$ may also be derived from physical dimensional arguments.

By differentiating both sides of Eq.\ref{nubar} by $\eta$ and then setting $\eta=1$, one obtains that
\begin{equation} \label{nuprime}
\ovl{\nu}\, ' (c) = \frac{\beta}{c} \, \ovl{\nu}(c),
\end{equation}
where $\beta \doteq g'(1) > 0$, from which it follows that
\begin{equation} \label{nupower}
\ovl{\nu}(c) \propto c^{\beta}.
\end{equation}
Changing integration variable in Eq.\ref{nuint} by $x \mapsto y= c \, x$, then differentiating both sides by $c$, and finally making use of Eq.\ref{nuprime}, some calculus leads to 
\begin{equation}
\int_{x_{0}}^{+\infty} \!\! dy \, f_{_{\!K}}\!\!\left(\frac{y}{c}\right) \nu(y) \left[ 1+\beta + \frac{y}{c} 
\frac{f'_{_{\!K}}\!\!\left(\frac{y}{c}\right)}{f_{_{\!K}}\!\!\left(\frac{y}{c}\right)} \right] = 0, \quad \forall c \in \mathbb{R}^{+},
\end{equation}
where $x_{0}=c \, \widehat{K}(c)$, which does not depend on $c$ because of the third postulate. It follows that
\begin{equation}
K \ \frac{f'_{_{\!K}}\!(K)}{f_{_{\!K}}\!(K)}  = -1-\beta,
\end{equation}
from which
\begin{equation}
f_{_{\!K}}\!(K) \propto K^{-\alpha},
\end{equation}
where $\alpha \doteq 1+\beta$, which is a further testable prediction, as the PDF of the sensitivity and that of the mean FR can be separately estimated empirically.

\section{Comparison with experimental data}

\subsection{Source of the experimental data}
To test the predicted dependence of mean FR on odorant concentration, experimental data was taken from Table S2 of \citet{Carlson2006}, a study on the \emph{Drosophila Melanogaster} olfaction\footnote{Because of the ease of access and of the amenability to genetic manipulation, electrophysiology and fluorescence imaging of fairly large populations of neurons, the \emph{Drosophila}'s has become the best known and studied animal model for olfactory systems.}, which provides the FRs of 24 ORNs exposed to 10 odorants, each at 4 different concentrations, one combination at a time (dimensionless dilution values: $10^{-8}$, $10^{-6}$, $10^{-4}$, $10^{-2}$), measured from counting the number of spikes within 500ms after stimulus delivery and averaging over 4-to-6 trials. The actual mean FRs are then plotted in Fig.\ref{analysis}a, where each point-symbol refers to one of the 10 odorants at the 4 different concentrations, on a log--log scale\footnote{The Table of \citet{Carlson2006} presents some negative values because the mean spontaneous activity was subtracted from the after-stimulus FRs, and some odorants had inhibitory effect on some ORNs. As ORNs that do not fire at all, because inhibited or just not excited enough by the odorant, obviously do not contribute to the actual stimulus-related mean FR of the population, which is the object of this work, all negative values in the Table are here reset to zero.}; the seemingly linear trend is already suggestive of a power-law relationship. 

\subsection{Probabilistic model for data analysis}
On the basis of a preliminary survey of the dataset, a probabilistic model is here proposed that takes into account several sources of variability and allows for a more reliable analysis of the data and, hence, for extracting more accurate information from them. The statistical validity of such model will be duly tested against the data. As it will appear more clearly in the following, one of the advantages of building this model for the experimental data is that, at some stage, several subsets of data-points can be pooled together, hence, in a way, providing a larger dataset for statistical inference.

The firing-rate $\mbf{Y}$ of any ORN is modelled with the function 
\begin{equation}\label{Pmodel}
\mbf{Y}=\mbf{A} \, \mrm{e}^{\sss\mbf{W}} c^{\beta} + \mbf{\cal{E}},
\end{equation}
where: the positive random variable (RV) $\bf{A}$ is determined by the choice of odorant; $\mbf{W}$ is a Gaussian RV of mean zero, determined by the choice of the pair concentration-odorant; $\mbf{\cal{E}}$ is a zero-mean noise that originates from the fluctuations of spike-counts between trials. The RVs $\mbf{A}$, $\mbf{W}$ and $\mbf{\cal{E}}$ are independent from each other. The RV $\mbf{A}$ is constant across measurements that use the same odorant as stimulus and is, in a way, equivalent to a \emph{quenched} noise, while $\mbf{W}$ is constant across trials that use the same odorant at the same concentration, and $\mbf{\cal{E}}$ is independently variable between measurements with any odorant at any concentration. The assumption that the standard deviation (s.d.) of $\mbf{\cal{E}}$ be independent from concentration and odorant is consistent with data presented in figure n. 3 of \citet{Carlson2006}, where the range of concentration covers four orders of magnitude for 10 odorants and 4 ORNs and where the s.d. also appears to be generally small in comparison to the FR. 

\begin{figure*}
\mbox{
\includegraphics[width=9cm,height=7cm]{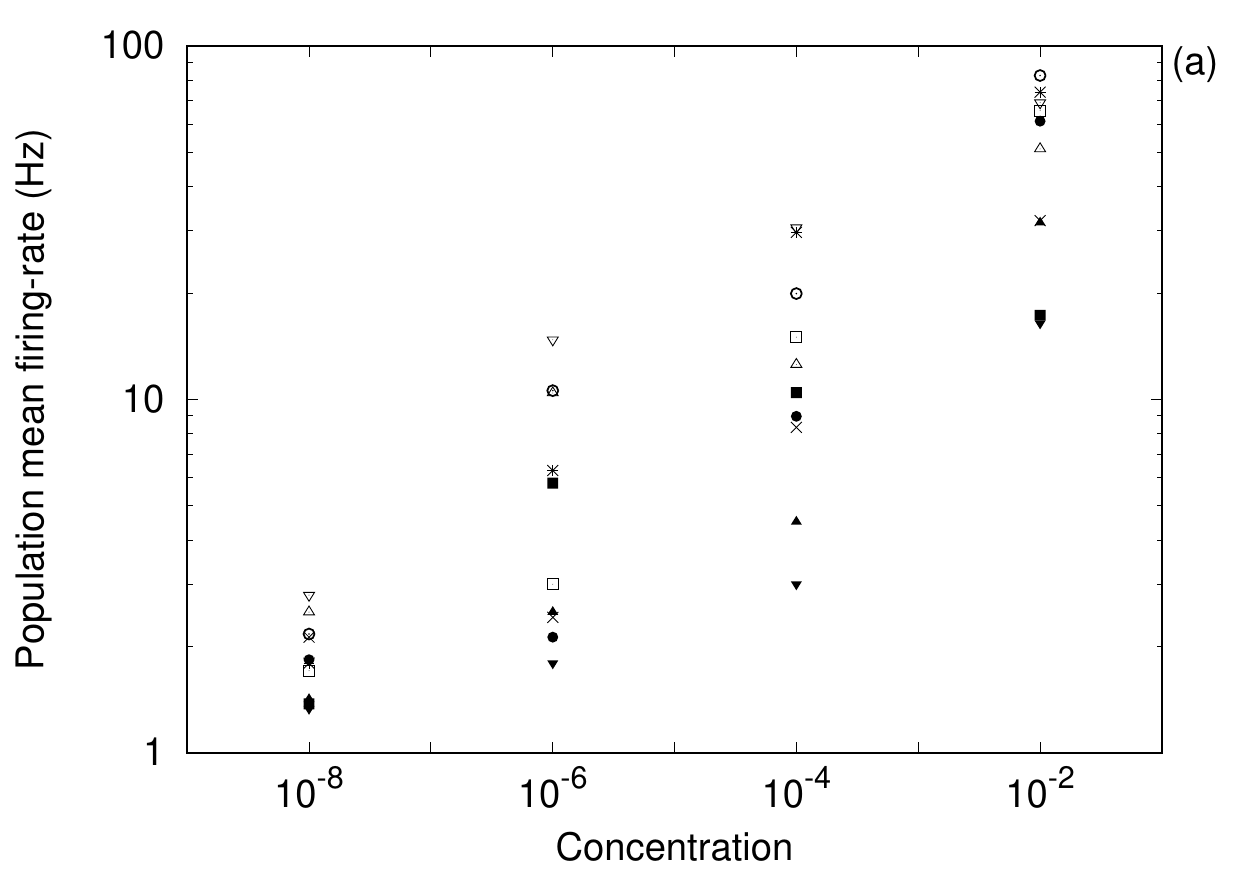}
\hspace{.2cm}
\includegraphics[width=9cm,height=7cm]{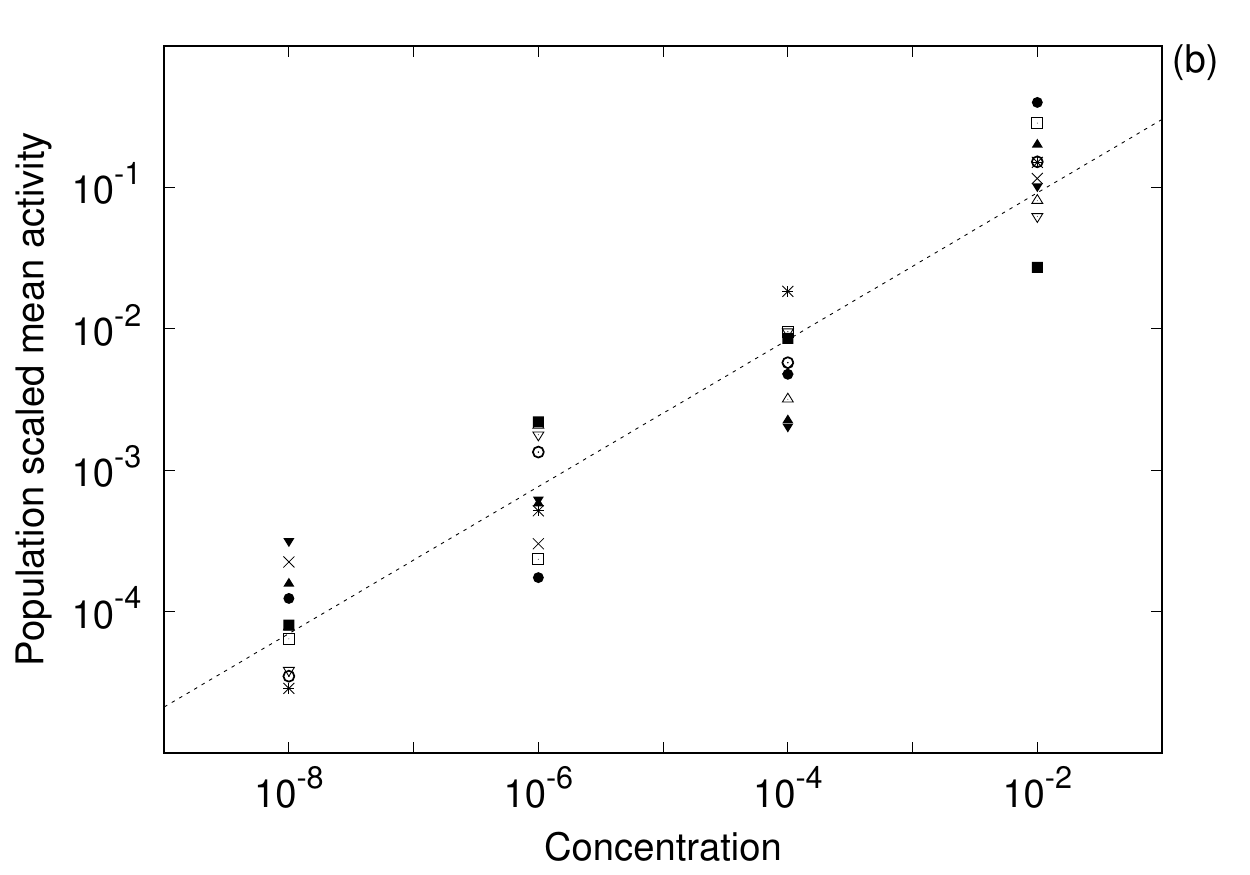} 
}
\caption{Plots of relevance to the data analysis: (a) Raw data plot of population mean firing-rate ($\mathrm{Hz}$) vs. odorants dilution (dimensionless), separately for each odorant (same symbol for the same odorant; $\log_{10}$ scale); (b) Firing-rates rescaled by the best estimate of the quenched noise and best-fit straight line, the data from all odorants having being pooled (slope $\simeq$ 0.2256).
\label{analysis}}
\end{figure*}

Both sides of Eq.\ref{Pmodel} are formally rescaled by a constant equal to 1 Hz, so that the quantities involved as well as the experimental figures become all dimensionless, maintaining the same symbols. Taking the logarithm of both sides of the equation, one obtains
\begin{equation} \label{lnfr}
\ln \mbf{Y}= \beta \ln c + \ln \mbf{A} + \mbf{W} + 
\ln \left(1 + \frac{\mbf{\cal{E}} \, c^{-\beta}}{\mbf{A} \, \mrm{e}^{\sss\mbf{W}}} \right).
\end{equation}
The term with noise $\mbf{\cal{E}}$ in the RHS also involves a dependence on $c$, which implies that the variance of $\ln(\mbf{Y})$ with respect to the joint PDF of all the random variables depends on the concentration. However, such dependence results to be statistically negligible, as also appears visually from Fig.\ref{analysis}a: for the lowest and the highest concentrations, there is some boundary effect (accumulation towards the extremes of activity, namely, quiescence and saturation respectively), but the variances in the two middle concentrations are statistically equivalent and similar in magnitude to the other two, despite the range of concentrations covering four orders of magnitude. It follows that the contribution of noise $\mbf{\cal{E}}$ is of minor relevance; in fact, this should not come as a surprise, for the noise term $\mbf{\cal{E}}$ is, in a way, a remnant of fluctuations already averaged over (same odorant, same concentration) trials and then ORNs. Accordingly with this observation, the last term in Eq.\ref{lnfr} will be ignored from this point onwards; the final verification of the model will support such approximation as statistically legitimate. 

Each realisation of the RVs, after self-explanatory renaming, is written as $y_{i j} = \beta \, x_{i} + a_{j} + W_{i j}$, where $i$ and $j$ run on the sets of $C$ concentrations and $D$ odorants, respectively. In the following, for convenience of notation, the unknown parameters $\left(a_{j}\right)_{j}$ will be considered as the components of the $D$-column $\ul{a}$, the controlled parameters $\left(x_{i}\right)_{i}$ as the components of the $C$-column $\ul{x}$, and the measured values $\left(y_{i j}\right)_{i j}$ as the components of the $C \! \times \! D$-matrix $\ul{y}$.

\subsection{Estimate of parameters}
The values in $\ul{y}$ and $\ul{x}$ are provided by the experiments, while the parameters to estimate are $\beta$, $\ul{a}$, and the variance $\sigma^{\sss{2}}$ of the Gaussian $\mbf{W}$. In order to lighten the notation, the controlled experimental variable $\ul{x}$ will be tacitly considered as given without uncertainty. Therefore, given the dataset $\ul{y}$, the probability density that the true values of the unknown parameters be $\sigma^{\sss{2}}$, $\beta$ and $\ul{a}$ is, by the Bayes formula,
\begin{equation} \label{condppar}
p(\sigma^{\sss{2}}, \beta, \ul{a} \, \vert \, \ul{y}) = p(\ul{y} \, \vert \, \sigma^{\sss{2}}, \beta, \ul{a}) \; p(\sigma^{\sss{2}}, \beta, \ul{a}) \, \slash \, p(\ul{y}),
\end{equation}
where $p(\sigma^{\sss{2}}, \beta, \ul{a})$ is the prior for the unknowns. In order to find the parameters values that are most likely to be the true ones given the empirical data, one can look for the values that maximise $p(\sigma^{\sss{2}}, \beta, \ul{a} \, \vert \, \ul{y})$ or, more conveniently, its logarithm. If one ignores the prior $p(\sigma^{\sss{2}}, \beta, \ul{a})$, this approach coincides with the standard Maximum Likelihood Method. Ignoring the prior is clearly equivalent to assuming arbitrarily that the prior is a positive constant in a sufficiently wide range and zero outside of it, as if that was the best way to mathematically represent null bias. In general, when the number of independent data-points is large in comparison to the number of unknowns, the contribution of the prior to the estimate is usually negligible, as appears when taking the logarithm. In the present case, the number of parameters to estimate is $D + 2 = 12$ and the number of data-points, once pooled together, is $C \cdot D = 48$; therefore, being these figures of similar magnitude, it would be inappropriate to ignore the prior. Null bias on the prior is realised here as a uniform distribution over a (sub-)space of the PDFs rather than on that of the parameters; this idea belongs with the theory of Information Geometry \citep{Amaribook}, built on the seminal works of \citet{Cencov}, \citet{Rao}, \citet{Hotelling}, \citet{Jeffreys1946}, and \citet{Fishermatrix}. Using the parameters as coordinates on a probability manifold and considering the submanifold spanned by the parameters ranges, giving equal weight to all volume elements of the submanifold translates generally into a non-uniform weighing of volume elements of the parameters space, the weight being given by $\sqrt{\mrm{det} \, g}$, where $g$ is the Riemannian metric on the manifold. It should be stressed that this approach is not entirely exempt from arbitrariness, for only a submanifold is taken into consideration rather than the whole space of PDFs, but is nevertheless much more representative of null-bias than the assumption of uniform PDF on the space of parameters. It was proved by \citet{Cencov} and later, differently and more generally, by \citet{Campbell} that, under what are essentially just consistency requirements, the only metric suitable for a probability manifold must be, but for an inconsequential arbitrary scaling constant, equal to
\begin{align}
g_{u v} =& \int d\ul{y}\; p(\ul{y} \, \vert \, \sigma^{\sss{2}}, \beta, \ul{a}) \,\cdot \\
& \cdot \left[ \partial_{u} \ln p(\ul{y} \, \vert \, \sigma^{\sss{2}}, \beta, \ul{a}) \right] \cdot \left[ \partial_{v} \ln p(\ul{y} \, \vert \, \sigma^{\sss{2}}, \beta, \ul{a}) \right],   \nonumber
\end{align}
known as the Rao-Fisher metric, where $u$ and $v$ are any two of the parameters, which in this case are $\sigma^{\sss{2}}$, $\beta$ and all the components of $\ul{a}$. Because the outcomes of the RV $\mbf{W}$ are independent across concentration-odorant pairs, one has that
\begin{equation}
p(\ul{y} \, \vert \, \sigma^{\sss{2}}, \beta, \ul{a}) = 
\prod_{i, j} \frac{1}{\sqrt{2 \pi \sigma^{\sss{2}}}} \, \exp\left\{- \frac{\left(y_{i j}- \beta x_{i} - a_{j}\right)^{2}}{2 \sigma^{\sss{2}}}\right\},
\end{equation}
where $i$ and $j$ run, as in previous expressions, over the $C$ concentrations and the $D$ odorants respectively. From this equation, one can calculate the components of the metric $g$ and hence its determinant, resulting in $\sqrt{\mrm{det} \, g} \propto \sigma^{\sss{-(D + 3)}}$, where, again, the proportionality coefficient is an inconsequential positive constant and $D$ is the number of odorants, that is, $D+3$ is the number of parameters to estimate plus one. The values of the parameters that maximise the logarithm of the LHS of Eq.\ref{condppar} are given by:
\begin{equation}
\left\{
\begin{array}{rcl}
\widehat{\beta} & = & \ds{\frac{\ovl{x y} - \ovl{x} \: \ovl{y}}{\ovl{x^{2}} - {\ovl{x}}^{2}}} \\
\\
\widehat{a}_{j} & = & \ds{\frac{1}{C} \sum_{i} y_{i j} - \widehat{\beta} \: \ovl{x}} \\
\\
\widehat{\, \sigma^{\sss{2}}} & = & \ds{\frac{1}{C D + D +3} \sum_{i, j} \left(y_{i j} - \widehat{\beta} \, x_{i} - \widehat{a}_{j} \right)^{2}}
\end{array}
\right.
\end{equation}
where the caps indicate the estimates and the over-bar indicates the arithmetic mean over all the $C \cdot D$ data-points (e.g., $\ovl{y} = \sum_{i, j} y_{i j} / (C D)$). The estimates of $\beta$ and of $\ul{a}$ result to be unbiased. The estimate of $\sigma^{\sss{2}}$, instead, is biased, which is not uncommon in maximisation estimations; in order to become unbiased, it only has to be multiplied by the factor $(C D + D + 3)/(C D - D -1)$. The eventual estimates are then: $\widehat{\beta} \simeq 0.2256$ and $\widehat{\, \sigma^{\sss{2}}} \simeq 0.1719$. As it turns out, having kept into consideration the prior is eventually made obsolete by rescaling to achieve unbiasedness, though it is not trivial at all that such correction, even after the conceptually important inclusion of the prior, is still only a constant prefactor.
\begin{figure*} 
\begin{center}
\includegraphics[]{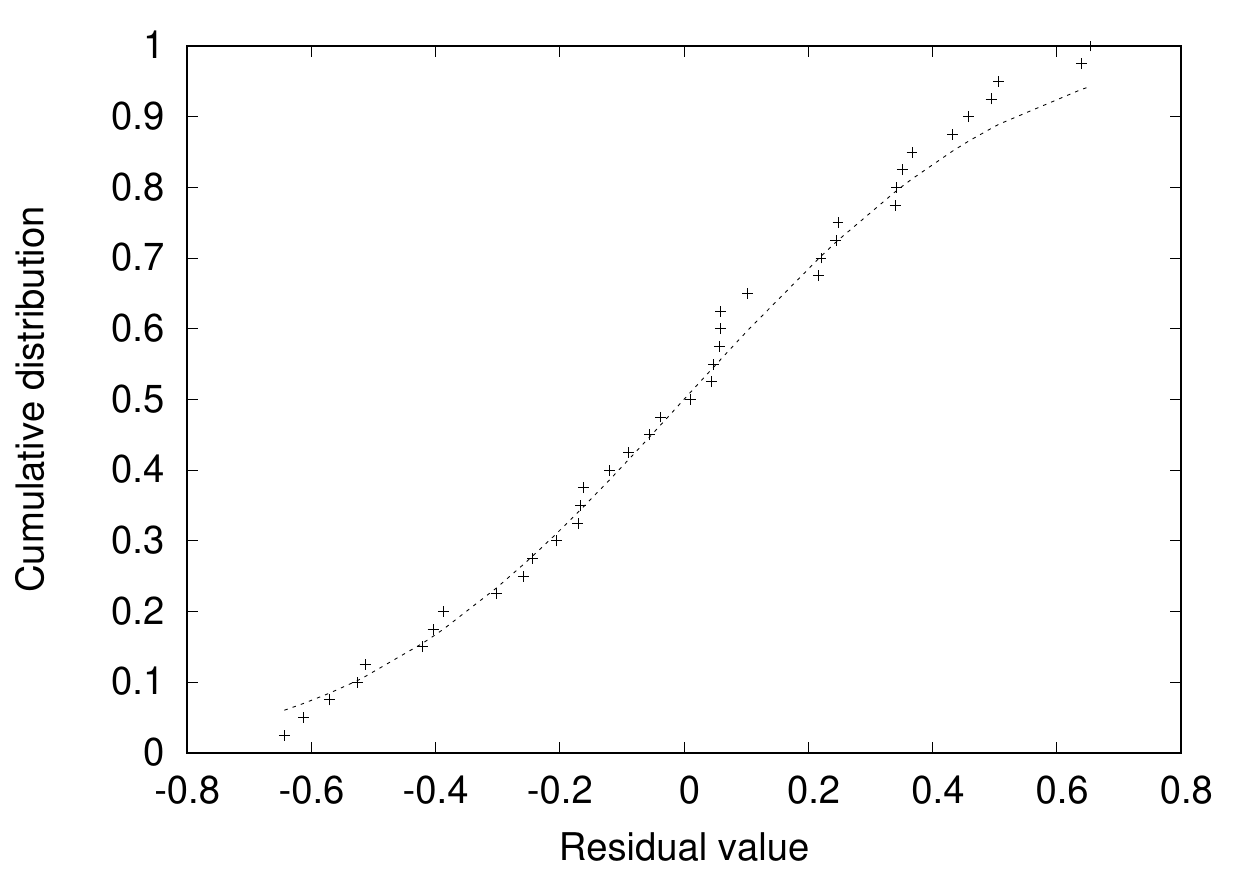}
\end{center}
\caption{Empirical (crosses) and model (dashed line) cumulative distributions of residuals (all residuals pooled together; $p<0.015$).
\label{fit}}
\end{figure*}
Having estimated $\ul{a}$, it is possible now to derive from the original dataset an odorant-independent (pseudo-)sample that pools together all of its $C \cdot D = 40$ data-points by rescaling the FRs, that is, $\left\{y_{i j} - \widehat{a}_{j}, \, \forall i, j \right\}$. The relative plot vs. the odorant concentration is shown in Fig.\ref{analysis}b together with the best-fit straight line. The \emph{empirical} cumulative distribution function (eCDF) of the residuals is
\begin{\eq}
\widehat{F}_{N}(t) \doteq \frac{1}{N} \sum_{n=1}^{N} T\left(r_{n} \leq t\right), \quad \forall t \in \mathbb{R}, \quad
\end{\eq}
where $T$ is the logical truth function, which is equal to 1 if the argument is true and to 0 if it is not, $N \doteq C D$, and $(r_{n})_{n}$ are the residual values. The Glivenko-Cantelli theorem guarantees that the eCDF converges uniformly to the theoretical CDF almost surely in the limit of infinite i.i.d. data-points. The eCDF is plotted against the best-fit Gaussian CDF in Fig.\ref{fit}; the apparent goodness of the fit will be quantified by means of a distribution-independent test.

\subsection{Goodness-of-fit test}
The plausibility of the model is evaluated using the eCDF; the first, immediate advantage of this approach is that it does not require `binning', whose intrinsic arbitrariness affects many other statistical tests. To quantify how well the model obtained in the previous Sections fits the data, use is made here of the goodness-of-fit (GOF) Smirnov-Cram\'er-von Mises (SCvM) test:
\begin{\eq} \label{scvmtest}
\omega^{2}_{N} \doteq N \int dF(t) \left[ \widehat{F}_{N}(t) - F(t) \right]^{2}.
\end{\eq}
The SCvM test belongs in the distribution-independent Smirnov subfamily \citep{Smirnov1937} of the Cram\'er-von Mises tests \citep{Cramer,vonMises2}. That the test is distribution-independent is made even more evident by the algebraic explicitation:
\begin{equation} \label{scvmexp}
\omega^{2}_{N} = \sum_{n}^{(z_{n})_{n}{\uparrow}} \left[ F(z_{n}) - \frac{2 n -1}{2 N} \right]^{2} + \frac{1}{12 N},
\end{equation}
where the up-arrow specifies that the terms in the sequence of data-points $(z_{n})_{n}$ are in ascending order; indeed, assuming that the data-points are from independent trials, the PDF of $F$ is uniform as long as $F$ is a continuous CDF. Therefore, the PDF of the test $\omega^{2}_{N}$ itself only depends on the number of data-points. In the numerical evaluation of the test variable, $F$ is a Gaussian CDF with mean zero and variance $\widehat{\, \sigma^{\sss{2}}}$. Although the asymptotic distribution of the SCvM test is known to be the integral of the square of the canonical Brownian Bridge stochastic process and can be expressed explicitly as a series \citep{Smirnov1937,Doob1949,Kac1949}, the small size of the dataset makes relying on such analytical derivation incautious. Therefore, an evaluation of the distribution of the test-variable was made through a Monte Carlo simulation with one million realisations over a set of the appropriate size ($N=40$; the pseudo-random number generation algorithm was adapted from routine \emph{ran2.c} of \citet{NR}, with double-precision arithmetic)\footnote{In order to eliminate the need for sorting routines, a less compact but equivalent algebraic explicit expression for $\omega^{2}_{N}$ can be easily derived from Eq.\ref{scvmtest} without assuming that the data values are ordered. Alternatively, use can be made of the \emph{ziggurat} algorithm of \citet{Marsaglia2004}, which directly generates ordered sets of uniform variates. One could also easily modify the formula to keep into account possible multiplicity of any data value, though with $N=40$ and 64-bit precision arithmetic this possibility can be safely ignored.}. The Monte Carlo results show that the probability of the SCvM test-variable taking any value below 0.02739 (critical value) by mere chance is 1.5\%, while the critical value for 1.0\% is 0.02522; the value obtained using Eq.\ref{scvmexp} is $\omega^{2}_{40} \simeq 0.02695$, corresponding to $p \simeq 0.0139$, therefore setting the estimate at $p < 0.015$.

\section{Conclusions}

\subsection{Summary and Results}
Biological olfactory systems are generally capable of maintaining the same odorant identification across a wide range of its concentration. This work was based on the following general hypotheses: (1) The olfactory systems compensate for changes of concentrations by normalisation; (2) The sensitivity $K$ of any ORN and the concentration $c$ of the presented odorant only enter the equations of the response dynamics through the product $K c$; (3) The sensitivity is statistically broadly distributed over ORNs and odorants and the statistical dependence between sensitivities is of minor relevance when considering statistics of large sets of ORNs and odorants.

Hypothesis (1) was translated into a mathematical relationship stating that the scaling of the concentration of the odorant leads to a corresponding scaling of the mean FR of the population of ORNs. Together with hypotheses (2) and (3), this lead mathematically to three predictions: (a) The population mean FR is a power-law function of the odorant concentration; (b) The PDF of the sensitivity across the population obeys a power-law; (c) The exponent of the power-law of the sensitivity is equal to that of the power-law of the concentration dependence plus one. All hypotheses and predictions have to be intended to be valid over a wide range of the relative variables but not over the whole range, as boundary and saturation effects may intervene.

Prediction (a) was tested against publicly available experimental data. In order to accurately analyse the data, a probabilistic model was introduced which takes into account separately three possible sources of randomness: odor\-ant-dependent finite-size effect, multiplicative noise, and trial-by-trial spike-count variability. The values of the parameters were estimated by means of info-geometric Bayesian maximisation followed by prefactor adjustment for unbiasedness, and the GOF was quantified by means of a finite-size distribution-independent SCvM test. The estimated model results to have $p<0.015$, with the estimate of the exponent of the power-law on concentration being $\widehat{\beta} \simeq 0.23$, which leads to the prediction for the exponent of the sensitivity power-law $\widehat{\alpha} \simeq 1.23$. Experimental data are also in agreement with predictions (b) and (c), but their support is weaker than for prediction (a) because of the large uncertainties in the datasets and relative extrapolations. Specifically, \citet{Samuel2019} extrapolated from their data sensitivity values that they found to follow a power-law PDF with exponent equal to about 1.42, but the relative statistical uncertainty is quite large\footnote{The authors adopted an unusual $p$-value, defined as the probability of obtaining a fictitious `empirical' dataset whose fitness be by mere chance further away from their best-fit model than the actual dataset, such probability resulting to be equal to about 0.17. The roots of this significant uncertainty appear to largely reside in the uncertainties of the best-fit interpolations of response curves. In particular, the majority of the interpolated datasets do not cover the neuronal response range up to saturation, therefore complicating the analysis and introducing even larger uncertainties. The authors also made use of a statistical method that eventually led them to exclude about half of their sensitivity data-points.} and, additionally, they experimented on the \emph{Drosophila} larva instead of the adult insect. 

\subsection{Discussion}
The physical interpretation of the random variable $\bf{A}$ is that of a finite-size effect: the number of ORs is quite small and it is unsurprising that the mean FR depends on the choice of odorant, while the population model predicts this dependence to become negligible for large numbers of ORs and ORNs. 

The multiplicative noise $\mrm{e}^{\sss\mbf{W}}$ has a log-normal PDF and may be interpreted physically as an approximation of the compound contribution of a number of positive random factors of comparable variances, accordingly with the Lindeberg theorem. The mean of $\bf{W}$ was chosen to be zero because the most likely outcome of the multiplicative noise is expected to be such to not affect the true value of the measured quantity; however, including a non-zero mean would not affect the results even if it depended on the choice of odorant, for it would be equivalent to an inconsequential scaling of $\bf{A}$. 

The exponent of the dependence of mean FR on odorant concentration is significantly smaller than the unit ($\widehat{\beta} \simeq 0.23$) with high statistical confidence, which may seem at odds with the hypothesis of divisive normalisation. However, while the results of \citet{Olsen2010} indicate that the inter-glomerular inhibition is an approximately linear function of the ORNs mean FR, they also indicate that such inhibition enters the response function of the relay neurons non-linearly. Therefore, the fact that the ORNs mean FR scales sub-linearly with the odorant concentration does not preclude normalisation as a plausible means to achieve concentration invariance.

The present model is statistical and is therefore likely more accurate for the mammalian olfactory system. Indeed, usually mammals have a much larger repertoire of receptor types as well as a much larger number of ORNs that express them than insects. These larger numbers should indeed make finite-size effects smaller, but they may possibly make the assumptions of this model be more closely fulfilled also because of a further reason: the larger numbers of receptor types and ORNs may have left the system less affected by evolutionary optimisation bias (including the incidence of inhibitory responses) towards the odorants most important for the survival of the species.

It seems worth noting that, if the subjective perception of the intensity (concentration) of an odorant is proportional to the mean FR of the population of ORNs, the predicted power-law dependence of mean population FR on odorant concentration is in agreement with the renowned psychophysical Stevens Law \citep{Stevens1957}: \emph{Equal stimulus ratios produce equal subjective ratios}, that is, in mathematical terms, the perceived intensity of a stimulus is proportional to a power of the actual physical intensity, at least over a relatively large range of the latter.

All in all, the approach adopted in this work shows how a wide variety of empirical evidence, ranging from molecular to psychophysical, may in fact be modelled starting from very few principles. The need for further experimental scrutiny of these principles might in itself suggest fruitful lines of study.

\subsection{Open questions}
The source of variability of $\bf{W}$ seems likely to lie in the experimental settings and procedures, probably in the control on the delivery of the odorant, whose concentration may vary because of uncertainties inherent in the apparatus. However, it cannot be entirely ruled out that there be at least a partial contribution from the studied system itself. In any case, the results suggest that such randomness arise from multiplicative sources. It would be of course more interesting if these fluctuations resulted not to be entirely an artifact, as physical processes responsible for the multiplicative randomness would have to be sought within the neural system.

It is notable that, were the sensitivity power-law coefficient $\alpha$ equal to 1, as it would be if the neuronal input followed the MAL of the simplistic ligand-receptor model $L+R \rightleftarrows LR$, the ORN population mean FR would be independent of the concentration ($\beta = 0$) but possibly for pseudo-random finite-size effects. Therefore, it is plausible that natural evolution has led to $\alpha$ being significantly different from 1 so that the population mean FR carry reliable information on concentration. However, while this observation may be of some interest in itself, it does not explain why, apart from being different from 1, the value of $\alpha$ is what it is. It seems likely that such value resulted from some kind of optimisation, but it seems equally likely that more understanding in this direction will require taking into account also further stages of the processing stream and their intrinsic interactions.

While the already available data seem sufficient to confirm the power-law dependence of ORNs mean FR on odorant concentration, at least in some animals, they are not enough to test with high accuracy the predictions on the sensitivity power-law and on the relationship between the power-laws respective exponents. More data are certainly needed, especially from experiments in which the sensitivities and the FRs are measured on the same animal model. This holds also true for a tighter verification of the postulate according to which the statistical dependence between affinities is small enough to be treated, to a good approximation on a large scale, as negligible ``short-range'' correlation. As already mentioned, the present modelling approach should be expected to be better suited to the mammalian olfactory system especially because of its much larger number of receptor types.

\section*{Competing interests}
No competing interests.

\section*{Acknowledgments}
All computer codes were compiled, using \emph{GCC} (\emph{GNU Compiler Collection}, Free Software Foundation, Inc.), and run on the open-source Linux-kernel \emph{Debian} OS.

\end{document}